\documentclass
[superscriptaddress,secnumarabic,amssymb,amsmath,nobibnotes,aps,prd,showkeys,showpacs,nofootinbib,onecolumn]{revtex4}%
\usepackage{graphicx}
\usepackage{epstopdf}
\usepackage{epsf}
\usepackage{bm}
\usepackage{amsmath}
\usepackage{amsfonts}
\usepackage{amssymb}
\usepackage{xcolor}%
\providecommand{\U}[1]{\protect\rule{.1in}{.1in}}

\newcommand{\be}{\begin{equation}}
\newcommand{\ee}{\end{equation}}

\newcommand{\mincir}{\raise
-3.truept\hbox{\rlap{\hbox{$\sim$}}\raise4.truept\hbox{$<$}\ }}
\newcommand{\magcir}{\raise
-3.truept\hbox{\rlap{\hbox{$\sim$}}\raise4.truept\hbox{$>$}\ }}

\begin{document}
\title{Dynamics and cosmological evolution in $\Lambda$-varying cosmology}
\author{G. Papagiannopoulos}
\email{yiannis.papayiannopoulos@gmail.com}
\affiliation{Faculty of Physics, Department of Astronomy-Astrophysics-Mechanics University
of Athens, Panepistemiopolis, Athens 157 83, Greece}
\author{Pavlina Tsiapi}
\affiliation{National Technical University of Athens, School of Applied Mathematical and
Physical Sciences, Iroon Polytechneiou 9, 15780, Athens, Greece}
\author{Spyros Basilakos}
\email{svasil@academyofathens.gr}
\affiliation{Academy of Athens, Research Center for Astronomy and Applied Mathematics,
Soranou Efesiou 4, 11527, Athens, Greece}
\affiliation{National Observatory of Athens, Lofos Nymphon - Thissio, PO Box 20048 - 11810,
Athens, Greece}
\author{Andronikos Paliathanasis}
\email{anpaliat@phys.uoa.gr}
\affiliation{Institute of Systems Science, Durban University of Technology, Durban 4000,
Republic of South Africa}

\begin{abstract}
We study the dynamical properties of a large body of varying vacuum
cosmologies for which dark matter interacts with vacuum. In particular,
performing the critical point analysis we investigate the existence and the
stability of cosmological solutions which describe de-Sitter, radiation and
matter dominated eras. We find several cases of varying vacuum models that
admit stable critical points, hence they can be used in describing the cosmic history.

\end{abstract}
\date{\today}
\maketitle

\section{Introduction}

The detailed analysis of the recent cosmological observations
\cite{dataacc1,dataacc2,data1,data2,data3,data4} indicates that in large
scales our universe is spatially flat and it consists of $\sim4\%$ baryonic
matter, $\sim26\%$ dark matter and $\sim70\%$ of dark energy (DE). Dark energy
is an \textquotedblleft exotic\textquotedblright\ fluid source with a negative
equation of state which attributes the cosmological acceleration. The origin
and nature of the DE is a complete mystery still, though some of its
properties are widely accepted, namely the fact that it has a negative
pressure. Obviously this has been a starting point that has given birth to
numerous alternative cosmological scenarios, which mainly generalize the
nominal Einstein-Hilbert action of General Relativity either by the addition
of extra fields \cite{hor1,hor2,hor3,hor4,ratra,peebles,tsujikawa1,hor6}, or a
non-standard gravity theory that increases the number of degrees of freedom
\cite{clifton,mod1,mod2,mod3,mod4,mod5,mod6}. These are two different
approaches in the dark energy problem which are still under debate in the
scientific community.

The introduction of a cosmological constant term, is one of the simplest ways
to modify the Einstein-Hilbert action. In the concordance $\Lambda$CDM model,
the cosmological constant coexists with the component of cold dark matter
(CDM) and baryonic matter. Although this model does describe the observed
universe quite accurately, it suffers from two basic problems, namely the
expected value of the vacuum energy density and the coincidence problem
\cite{conpr1,conpr2,conpr3,conpr4}. An interesting approach for solving those
problems is to allow $\Lambda$ to vary with cosmic time, see
\cite{Bas2009a,Bass1,Bass1a,perico,Bas09a,Tsiapi:2018she} and references
therein. These models
\cite{Ozer,bertolami,chenwu,lim00,lima6,lima7,many1,many2,many3,aldrovandi,Schutz,Schutz2,Lima1,Lima2,Lima4,Lima5,Waga}
are based on a dynamical $\Lambda$ term that evolves as a power series of the
Hubble rate \cite{ShapiroSola, Sola2,Sool14}. It was found that in the latter
models the spacetime can be the physical result of a non-singular initial de
Sitter vacuum stage, that also provides a graceful transition out of the
inflation and into the radiation era. It has been found that these running
vacuum scenarios accommodate the radiation and matter dominated era as well as
the late time cosmic acceleration \cite{Bass1,perico,bas22}.

In this context, matter is allowed to interact with dark energy
\cite{Tsiapi:2018she,Amendola1,Amendola2,DelCampo1,DelCampo2,Pavon,in1,in2,in3,in4,in5,Panotopoulos}%
. Although, this interaction is not imposed by a fundamental principle, it has
its roots in the particle physics theory, where any two matter fields can
interact with each other. Such an interaction has been found to be a very
efficient \ way to explain the cosmic coincidence problem and at the same time
approach the mismatched value of the Hubble constant $H_{0}$ from the global
$\Lambda$CDM based Planck and local measurements. Thus, in the present work we
shall consider several interacting cosmological models of\ $\Lambda\ $varying
cosmologies. The structure of the manuscript is as follows.

In Section 2, we briefly introduce the concept of the running $\Lambda
\ $varying cosmologies and the interacting models that we shall study. Section
3, includes the main analysis of our work where we study the dynamical
behaviour of our models and present the main results of this work. More
specifically we study the critical points and their stability. Each critical
point describes a specific exact solution for the field equation which
correspond to the cosmic history. By studying the stability of the solutions
of the critical points we are able to reconstruct the cosmic history and infer
about the cosmological viability of these models. Finally, in Section 4, we
summarize our results and we draw our conclusions.

\section{$\Lambda$-varying Cosmology}

We consider a universe with a perfect fluid with energy density $\rho$, and
pressure $p=w\rho$; such that the energy-momentum tensor is given by ${T}%
_{\mu\nu}=-p\,g_{\mu\nu}+(\rho+p)U_{\mu}U_{\nu}.$ In addition we consider the
$\Lambda-$varying cosmological term, $T_{\mu\nu}^{\left(  \Lambda\right)  }=$
$\rho_{\Lambda}\left(  t\right)  \,g_{\mu\nu}$,~$\rho_{\Lambda}=\Lambda\left(
t\right)  /(8\pi G)$ where the effective energy momentum tensor is written as
$\tilde{T}_{\mu\nu}\equiv T_{\mu\nu}+g_{\mu\nu}\rho_{\Lambda}$.

In General Relativity $\rho_{\Lambda}$ is considered to be constant; however
in varying vacuum cosmology, $\Lambda$ is considered to be a function of the
cosmic time, or of any collection of homogeneous and isotropic dynamical
variables, i.e. $\Lambda=\Lambda(\chi(t))$. \ 

The Einstein field equations are written as,%
\begin{equation}
R_{\mu\nu}-\frac{1}{2}g_{\mu\nu}R=8\pi G\ \tilde{T}_{\mu\nu}\,. \label{EE}%
\end{equation}
where on the lhs part is the Einstein tensor and on the rhs the effective
energy momentum tensor. For spatially flat FLRW spacetime with line element%
\begin{equation}
ds^{2}=-dt^{2}+a^{2}\left(  t\right)  \left(  dx^{2}+dy^{2}+dz^{2}\right)  ,
\end{equation}
the Friedmann equations are%
\begin{equation}
3H^{2}\left(  t\right)  =\Lambda\left(  t\right)  +\rho\left(  t\right)  ,
\label{fe.01}%
\end{equation}%
\begin{equation}
-2\dot{H}\left(  t\right)  -3H^{2}\left(  t\right)  =-\Lambda\left(  t\right)
+p\left(  t\right)  , \label{fe.02}%
\end{equation}
where we have set $8\pi G\equiv c\equiv1$ and $H\left(  t\right)  =\frac
{\dot{a}\left(  t\right)  }{a(t)}$ is the Hubble function.

In this work we shall consider a universe with radiation, dark and baryonic
pressureless matter as well as the varying $\Lambda~$term, hence the Friedmann
equations (\ref{fe.01}), (\ref{fe.02}) take the following form%

\begin{equation}
3H^{2}=(\rho_{m}+\rho_{r}+\rho_{\Lambda}), \label{FR1}%
\end{equation}

\begin{equation}
2\dot{H}+3H^{2}=-(\frac{1}{3}\rho_{r}-\rho_{\Lambda}), \label{FR2}%
\end{equation}
where we have used $\rho_{m}=\rho_{DM}+\rho_{b}.$ Assuming that baryons and
radiation are self-conserved, namely the corresponding densities evolve in the
nominal way, $\rho_{r}=\rho_{r0}a^{-4}$ for the radiation density and
$\rho_{b}=\rho_{b0}a^{-3}$ for the baryon density. In this way we only
consider interaction between the Dark Matter and the varying vacuum sectors.
Thus the Bianchi identity gives:%
\begin{equation}
\dot{\rho}_{_{DM}}+3H\rho_{_{DM}}=-\dot{\rho}_{_{\Lambda}}=Q, \label{FR3}%
\end{equation}
where $Q$\ is the interaction term between the Dark Matter and the varying
vacuum component, which we will study in this work in order to define their
dynamical behavior. Here we investigate the generic evolution of the solution
which is described by the field equations (\ref{FR1}), (\ref{FR2}) and
(\ref{FR3}) for specific functional forms of the interaction term $Q$.
Specifically, we shall consider five different cases:

The first case that we study is the \textit{running vacuum} model (RVM) (see
\cite{Bas2009a,Bass1,perico,Tsiapi:2018she}). Theoretical motivations for this
model arise from Quantum Field Dynamics (QFT) in curved space-time, by
associating Renormalization Group's running scale $\mu$ (in our context the
dynamical parameter $\chi(t)$) with a characteristic energy threshold for
cosmological scales. Thus, $\chi(t)$ is chosen to be the Hubble rate $H$, for
reviews see \cite{RG1,RG2,RG3}.

Returning to our definition of $\Lambda(t)=\Lambda(\chi(t))$, we may express
the running vacuum as a power series of the Hubble function:
\[
\Lambda(t)=\Lambda(H(t))=c_{0}+\sum_{k}\alpha_{k}H^{k}(t).
\]

It has been shown in previous works, that only even powers of $H$ can be
theoretically motivated, as the odd powers of the Hubble function are
incompatible with the general covariance of the effective action
\cite{Sola:2007sv,Shapiro:2009dh}. For that reason we shall exclude odd powers
of $H$ from the series. Furthermore, high powers of $H$ can be very useful
when treating the evolution of the early universe, but they are negligible in
the matter and dark energy eras respectively \cite{perico}. In this study we
are restricting our analysis to the simplified model
\cite{JSola,ShapiroSola,StarobinskyPhysLet,LHsingular,bassola}:%

\begin{equation}
\Lambda(H)=c_{0}+nH^{2},
\end{equation}
where $n$ is a dimensionless parameter, linked to the strength of the
interaction. For consistency, the condition $\rho_{\Lambda}(H_{0}%
)=\rho_{\Lambda0}=\frac{\Omega_{\Lambda}}{\rho_{crit}}$ fixes the value of
$c_{0}$ at $c_{0}=H_{0}^{2}(\Omega_{\Lambda}-n)~$\cite{Gomez-Valent:2014rxa}.
In the case of the RVM, the interaction term is taken by solving the
continuity equation (\ref{FR3}) for the specific form of $\rho_{\Lambda}%
=\frac{3}{8\pi G}\Lambda(H)=\rho_{\Lambda0}+\frac{3}{8\pi G}nH$, and is given by:%

\begin{equation}
Q_{A}=nH(3\rho_{DM}+3\rho_{b}+4\rho_{\Lambda}).
\end{equation}
In the second vacuum scenario used in this study the corresponding interaction
term is taken ad hoc to be proportional to the density of dark matter
\cite{Salvateli}. In particular, the interaction term is given by
$Q_{B}=3nH\rho_{DM}~$ where, as before, the dimensionless parameter $n$ is an
indicator of the interaction strength. Then we examine a third vacuum scenario
which is presented in \cite{Tsiapi:2018she,QproptorhoL} where the interaction
term is written as $Q_{C}=3nH\rho_{\Lambda}.~$ Motivated by interesting
results on the above models, we also considered two additional scenarios.

The fourth model of our study is $Q_{D}=\frac{3n}{H}\rho_{b}\rho_{DM}~$where
the interaction is dependent also on the baryonic density as well as dark
matter, while for the last model of our study we assume $Q_{E}=3nH\rho
_{tot},~$in which the total density affects the interaction term.

To this end, from the observational viewpoint the values of $n$ are found to
be quite small, pointing a small (but not zero) deviation from the usual
$\Lambda$CDM model. Indeed, the concordance model is recovered in any case
when $n$ is set to $0$. For the first three vacuum models, $n$ is treated as a
free parameter along with other cosmological parameters and it is found to be
of the order $\sim10^{-3}$ or less, see for example
\cite{Gomez-Valent:2014rxa,Sola:2016jky}, where these authors found
$n=0.00013\pm0.00018$, $n=0.00014\pm0.00103$). Interactions $Q_{A},~Q_{B},$
$Q_{C}$ and $Q_{E}$ can be seen as linear interaction terms while $Q_{D}$ is a
nonlinear function.

\section{Dynamical Analysis}

In this Section, we study the cosmological evolution of the aforementioned
cosmological scenarios by using methods\ of dynamical systems \cite{con1,con2}%
. Specifically we study the critical points of the field equations in order to
identify the cosmological eras that are provided by the theory. The stability
of those cosmological eras are determined by calculating the eigenvalues of
the linearized system at the critical point. The way we approach this analysis
is described as follows.

We define proper dimensionless variables to rewrite the field equations so
that our analysis can be universal. Then we proceed by producing the
first-order ordinary differential equations from our dimensionless variables.
The critical points of the system are those sets of variables for which every
differential equation of our system is equal to zero. These sets of variables
represent different epochs of the cosmos that we further study in order to
consider them as potential candidates that actually describe the observed
universe. The eigenvalues of those points are important tools towards
characterizing the stability of the critical points \cite{Liddle}.

If a critical point is stable/attractor then the corresponding eigenvalues
will need to have negative real parts. Thus, the eigenvalues can be used in
order to understand the behavior of the dynamical system around the critical
point \cite{wiggins}.\newline\qquad Our approach is as follows. We consider a
dynamical system of any number of equations:
\[
\dot{x}^{A}=f^{A}\left(  x^{B}\right)  ,
\]
then a critical point of the system, namely $P=P\left(  x^{B}\right)  $
satisfies \ $f^{A}\left(  P\right)  =0$. The linearized system around $P$ is
written as%
\[
\delta\dot{x}^{A}=J_{B}^{A}\delta x^{B},~J_{B}^{A}=\frac{\partial f^{A}\left(
P\right)  }{\partial x^{B}}.
\]
where $J_{B}^{A}$ is the respective Jacobian matrix. We calculate the
eigenvalues and eigenvectors and write the general solution on the respective
points as their expression. Since the linearized solutions are expressed in
terms of the eigenvalues $\lambda_{i}$ as functions of $e^{\lambda_{i}t}$,
when all those terms have negative real parts the solution on the critical
point is apparently stable.

\subsection{Dimensional system}

In order to study the generic evolution of the cosmological models of our
consideration we prefer to work in the $H-$normalization where define the
dimensionless variables \cite{con1,con2}%

\[%
\Omega
_{DM}=\frac{\rho_{_{DM}}}{3H^{2}},~%
\Omega
_{r}=\frac{\rho_{r}}{3H^{2}},~%
\Omega
_{b}=\frac{\rho_{b}}{3H^{2}},~%
\Omega
_{\Lambda}=\frac{\rho_{\Lambda}}{3H^{2}}%
\]
Consequently, the constraint equation (\ref{FR1}) becomes%
\begin{equation}%
\Omega
_{DM}+%
\Omega
_{r}+%
\Omega
_{b}+%
\Omega
_{\Lambda}=1, \label{con1}%
\end{equation}
while the rest of the field equations can be written as the following
four-dimensional first-order ordinary differential equations%

\begin{equation}
\frac{d%
\Omega
_{DM}}{d\ln a}=-%
\Omega
_{DM}(3+2\frac{\dot{H}}{H^{2}})-\frac{Q}{3H^{3}},
\end{equation}

\begin{equation}
\frac{d%
\Omega
_{r}}{d\ln a}=-2%
\Omega
_{r}(2+\frac{\dot{H}}{H^{2}}),
\end{equation}

\begin{equation}
\frac{d%
\Omega
_{b}}{d\ln a}=-2%
\Omega
_{b}(\frac{3}{2}+\frac{\dot{H}}{H^{2}}),
\end{equation}

\begin{equation}
\frac{d%
\Omega
_{\Lambda}}{d\ln a}=-2%
\Omega
_{\Lambda}\frac{\dot{H}}{H^{2}}-\frac{Q}{3H^{3}},
\end{equation}
in which
\begin{equation}
\frac{\dot{H}}{H^{2}}=\frac{1}{2}(3%
\Omega
_{\Lambda}-%
\Omega
_{r}-3).
\end{equation}
and as new independent variable we consider the number of e-fold $N=\ln a.$

By using the constraint equation (\ref{con1}) we are able to reduce the latter
dynamical system into the following three-dimensional system%

\begin{equation}
\frac{d%
\Omega
_{r}}{d\ln a}=-%
\Omega
_{r}(-1-3%
\Omega
_{\Lambda}+%
\Omega
_{r}), \label{ds1}%
\end{equation}

\begin{equation}
\frac{d%
\Omega
_{b}}{d\ln a}=-%
\Omega
_{b}(3%
\Omega
_{\Lambda}-%
\Omega
_{r}), \label{ds2}%
\end{equation}

\begin{equation}
\frac{d%
\Omega
_{\Lambda}}{d\ln a}=-%
\Omega
_{\Lambda}(3%
\Omega
_{\Lambda}-%
\Omega
_{r}-3)-\frac{Q}{3H^{3}} \label{ds3}%
\end{equation}
The latter equation depends on the functional form of $Q$, which is necessary
to be defined in order to continue with our analysis.

\subsection{Case A - $Q_{A}$}

For the first model of our consideration in which $Q_{A}=nH(3\rho_{DM}%
+3\rho_{b}+4\rho_{r})$, equation (\ref{ds3}) becomes%

\begin{equation}
\frac{d%
\Omega
_{\Lambda}}{d\ln a}=-%
\Omega
_{\Lambda}(3%
\Omega
_{\Lambda}-%
\Omega
_{r}-3)-n\ (3-3%
\Omega
_{\Lambda}+%
\Omega
_{r}), \label{ds3a}%
\end{equation}

Hence, by assuming the rhs of equations (\ref{ds1}), (\ref{ds2}), (\ref{ds3a})
to be zero we determine the critical points of the dynamical system. Every
point $P$ has coordinates $P=\mathbf{\{}%
\Omega
_{DM},%
\Omega
_{b},%
\Omega
_{\Lambda},%
\Omega
_{r}\ \mathbf{\}}$ and describes a specific cosmological solution. For every
point we determine the physical cosmological variables as also the equation of
the state parameter. In order to determine the stability of each critical
point the eigenvalues of the linearized system around the critical point $P$
are derived. \ Therefore, the dynamical system (\ref{ds1}), (\ref{ds2}),
(\ref{ds3a}) admits the three critical points with coordinates $A_{1}%
=\{0,0,1,0\},~B_{1}=\{-4n,0,n,1+3n\}$ and $C_{1}=\{1-n,0,n,0\}$

Point $A_{1}~$describes a de Sitter universe with equation of state
parameter~$w=-1$, where only the cosmological constant term contributes in the
evolution of the universe. The eigenvalues of the linearized system are found
to be $\{-4,-3,-3(1-n)\}$ from where we infer that point $A_{1}$ is an
attractor for $n<1$. This is in agreement with the expected values of $n$ and
thus this point is of physical interest.

Point $B_{1}$ is physical accepted when $\,-\frac{1}{4}\leq n\leq0.$ In this
area these points correspond to a universe where radiation, dark matter and
the cosmological constant coexist and dynamically it behaves like a radiation
dominated universe ($w=\frac{1}{3}$) which is the case for $n\rightarrow0$.
The eigenvalues of the linearized system at the point $B_{1}$ are derived to
be $\left\{  4,1,1+3n\right\}  $ from where we conclude that the point is a
source (unstable point).

\begin{figure}[ptb]
\centering\includegraphics[scale=0.6]{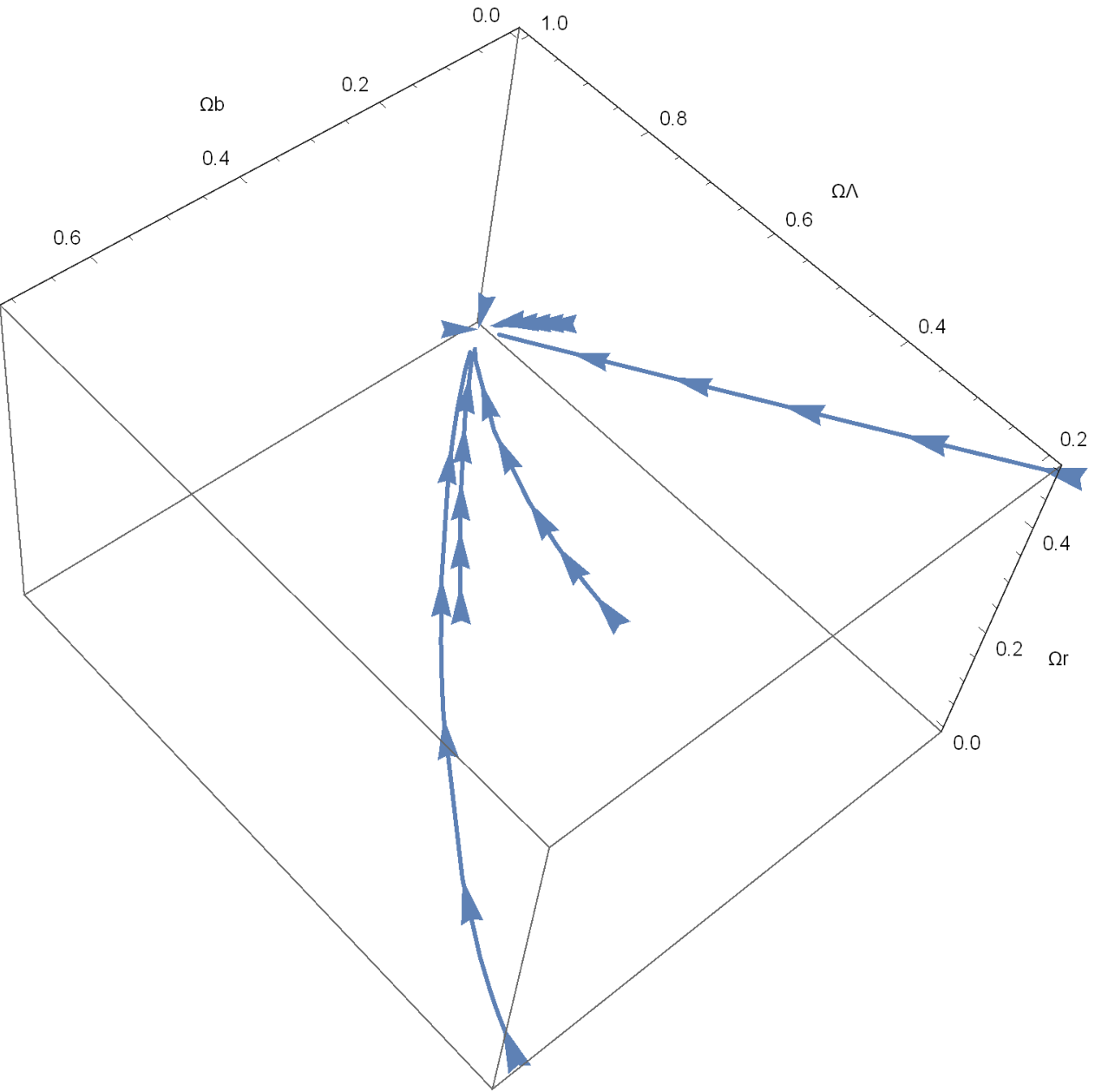}\caption{Phase space diagram
for the dynamical system (\ref{ds1}), (\ref{ds2}), (\ref{ds3a}).We consider
(a) $\Omega_{b}=0.2996,\Omega_{r}=0.0004$,$\Omega_{\Lambda}=0.7$ (b)
$\Omega_{b}=0,\Omega_{r}=0.1$,$\Omega_{\Lambda}=0.9$ (c) $\Omega_{b}%
=0.3$,$\Omega_{r}=0.2$,$\Omega_{\Lambda}=0.5$ (d) $\Omega_{b}=0$,$\Omega
_{r}=0.5$,$\Omega_{\Lambda}=0.2$ (e) $\Omega_{b}=0.7$,$\Omega_{r}=0.1$%
,$\Omega_{\Lambda}=0.2$, for $n<1$. The unique attractor is the de Sitter
point $A_{1}$}%
\label{fig1}%
\end{figure}

\begin{figure}[ptb]
\centering
\includegraphics[scale=0.6]{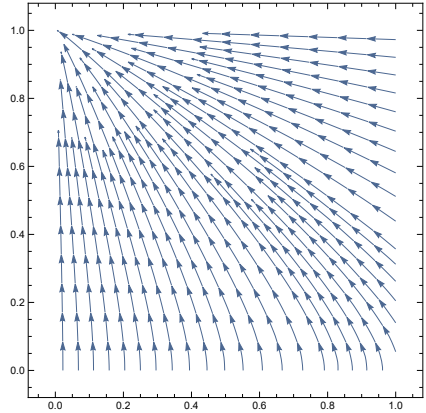}\caption{Phase space diagram for the
dynamical system (\ref{ds1}), (\ref{ds2}), (\ref{ds3a}) in the space of
variables $\Omega_{b},\Omega_{\Lambda}$ for $n<1$ and $\Omega_{r}=10^{-4}$.
The unique attractor is the de Sitter point $A_{1}$.}%
\label{fig2}%
\end{figure}

Point $C_{1}$ describes a universe where only the cosmological constant and
the dark matter fluids contribute to the total cosmic fluid. Indeed it
describes the $\Lambda-$CDM universe where now the parameter $n$ is the energy
density of the cosmological constant, i.e. $\Omega_{\Lambda}=n$. The point is
physical accepted when $0\leq n\leq1$, while for $n=1$ it is reduced to point
$A_{1}$. The eigenvalues of the linearized system are determined to be
$\{-1-3n,3\left(  1-n\right)  ,-3n\}$ from where we infer that the solution of
the critical point is always unstable. The critical point analysis of the
above system yields three critical points that are shown in Table
\ref{table1}. In Figs. \ref{fig1} and \ref{fig2} the phase space diagram of
the dynamical system $Q_{A}$ is presented for $n<1$ \ ($n=-0,1$ ) from where
we can see that the unique attractor is the de Sitter point $A_{1}$.%

\begin{table}[tbp] \centering
\caption{Critical points and physical quantities for Case A}%
\begin{tabular}
[c]{ccccccc}\hline\hline
$\text{Point}$ & $\{%
\Omega
_{DM},%
\Omega
_{b},%
\Omega
_{\Lambda},%
\Omega
_{r}\ \}$ & existence & $w$ & A$\text{cceleration}$ & E$\text{igenvalues}$ &
Stability\\\hline
$A_{1}$ & $\{0,0,1,0\}$ & Always & $-1$ & Yes & $\{-4,-3,-3(1-n)\}$ & Stable
for $n<1$\\
$B_{1}$ & $\{-4n,0,n,1+3n\}$ & $-\frac{1}{4}\leq n\leq0$ & $\frac{1}{3}$ &
No & $\left\{  4,1,1+3n\right\}  $ & Unstable\\
$C_{1}$ & $\{1-n,0,n,0\}$ & $0\leq n\leq1$ & $-n$ & Yes for $n>\frac{1}{3}$ &
$\{-1-3n,-3(n-1),-3n\}$ & Unstable\\\hline\hline
\end{tabular}
\label{table1}%
\end{table}%

\subsection{Case B - $Q=3nH\rho_{DM}$}

In this case our system of study are equations (\ref{ds1}), (\ref{ds2}) and%

\begin{equation}
\frac{d%
\Omega
_{\Lambda}}{d\ln a}=-%
\Omega
_{\Lambda}(3%
\Omega
_{\Lambda}-%
\Omega
_{r}-3)-3n(1-%
\Omega
_{b}-%
\Omega
_{\Lambda}-%
\Omega
_{r})\ , \label{ds3b}%
\end{equation}
thus the dynamical system (\ref{ds1}), (\ref{ds2}), (\ref{ds3b}) admits four
critical points with coordinates $A_{2}=\{0,0,1,0\},~B_{2}=\{\frac{3-n}%
{3},0,\frac{n}{3},0\}$ and $C_{2}=\{0,1,0,0\},D_{2}=\{0,0,0,1\}$

Point $A_{1}~$describes a de Sitter universe with an equation of state
parameter~$w=-1$, where only the cosmological constant term contributes in the
evolution of the universe. The eigenvalues of the linearized system are found
to be $\{-4,-3,-3+n\}$ and thus we can conclude that point $A_{2}$ is an
attractor for $n<3$. Taking into account the literature values of $n$
\cite{Gomez-Valent:2014rxa}, this is a valid point.

Point $B_{2}$ provides a $\Lambda$CDM\textbf{ }scenario where the components
of the fluid are $%
\Omega
_{DM}=1-\frac{n}{3}~$and $%
\Omega
_{\Lambda}=\frac{n}{3}\mathbf{.}$Aparently this family of points exists only
for $0\leq n\leq3,$ but it can be an accelerating point only for $n>1$. For
$n=3$ this point reduces to a deSitter one. In terms of stability, the
eigenvalues of the linearized system are $\{3-n,-n,-n-1\},$ hence, this point
is an attractor, i.e. stable for $n>3$, while it is a source for $n<3$.

Point $C_{2}$ describes a baryon dominated universe, while the solution at
this point is always unstable since there is always a positive eigenvalue,
namely the corresponding eigenvalues are $\{3,-1,n\}$.

Point $D_{2}$\ describes a radiation dominated universe that does not
accelerate, the corresponding eigenvalues are $\{4,1,1+n\},$ hence the current
point is a source.

The critical point analysis of the above system yields four critical points
that are shown in Table \ref{table2}%

\begin{table}[tbp] \centering
\caption{Critical points and physical quantities for Case B}%
\begin{tabular}
[c]{ccccccc}\hline\hline
$\text{Point}$ & $\{%
\Omega
_{DM},%
\Omega
_{b},%
\Omega
_{\Lambda},%
\Omega
_{r}\ \}$ & E$\text{xistence}$ & $w_{tot}$ & A$\text{cceleration}$ &
E$\text{igenvalues}$ & S$\text{tability}$\\\hline
$A_{2}$ & $\{0,0,1,0\}$ & Always & $-1$ & Yes & $\{-4,-3,-3+n\}$ & Stable for
$n<3$\\
$B_{2}$ & $\{\frac{3-n}{3},0,\frac{n}{3},0\}$ & $0\leq n\leq3$ & $-\frac{n}%
{3}$ & Yes for$~n>1$ & $\{3-n,-n,-n-1\}$ & Stable for$~n>3$\\
$C_{2}$ & $\{0,1,0,0\}$ & Always & $0$ & No & $\{3,-1,n\}$ & Unstable\\
$D_{2}$ & $\{0,0,0,1\}$ & Always & $\frac{1}{3}$ & No & $\{4,1,1+n\}$ &
Unstable\\\hline\hline
\end{tabular}
\label{table2}%
\end{table}%

\begin{figure}[ptb]
\centering\includegraphics[scale=0.6]{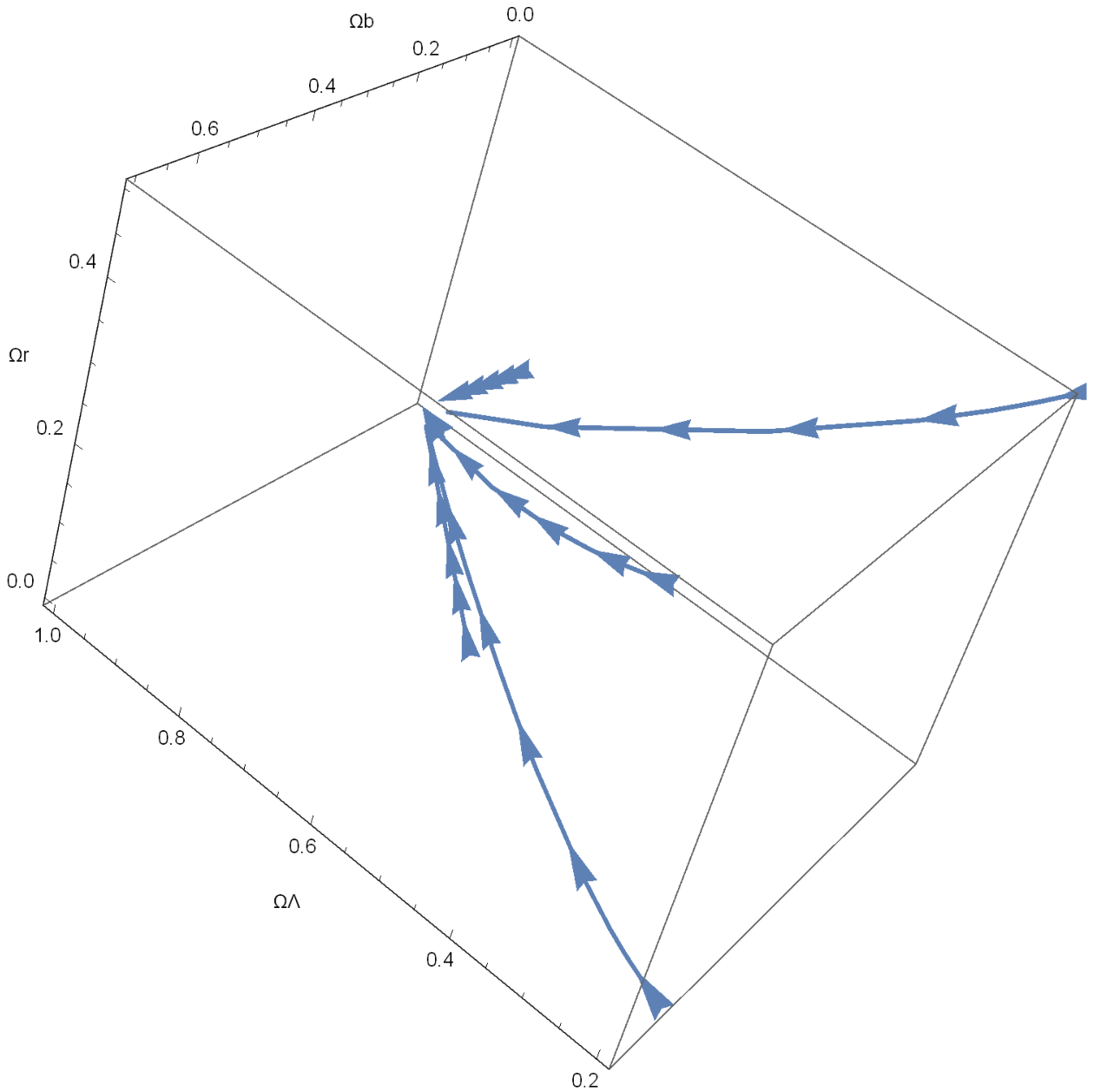}\caption{Phase space diagram
for the dynamical system (\ref{ds1}), (\ref{ds2}), (\ref{ds3b}).We consider
(a) $\Omega_{b}=0.2996,\Omega_{r}=0.0004$,$\Omega_{\Lambda}=0.7$ (b)
$\Omega_{b}=0,\Omega_{r}=0.1$,$\Omega_{\Lambda}=0.9$ (c) $\Omega_{b}%
=0.3$,$\Omega_{r}=0.2$,$\Omega_{\Lambda}=0.5$ (d) $\Omega_{b}=0$,$\Omega
_{r}=0.5$,$\Omega_{\Lambda}=0.2$ (e) $\Omega_{b}=0.7$,$\Omega_{r}=0.1$%
,$\Omega_{\Lambda}=0.2$, for $n<1$. The unique attractor is the de Sitter
point $A_{2}$}%
\label{fig3}%
\end{figure}

\begin{figure}[ptb]
\centering
\includegraphics[scale=0.6]{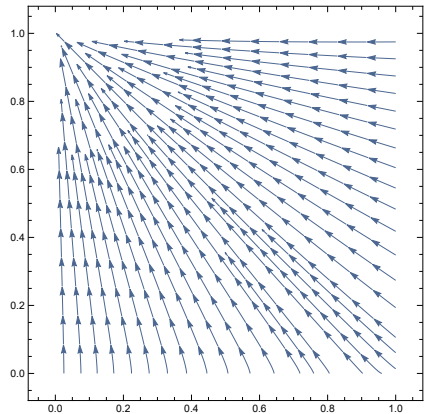}\caption{Phase space diagram for the
dynamical system (\ref{ds1}), (\ref{ds2}), (\ref{ds3b}) in the space of
variables $\Omega_{b},\Omega_{\Lambda}$ for $n<1$ and $\Omega_{r}=10^{-4}$.
The unique attractor is the de Sitter point $A_{2}$.}%
\label{fig4}%
\end{figure}

In Figs. \ref{fig3} and \ref{fig4} the phase space diagram of the dynamical
system $Q_{B}$ is presented for $n<1$ \ ($n=-0,1$ ) from where we can see that
the unique attractor is the de Sitter point $A_{2}$. 

\subsection{Case C - $Q=3nH\rho_{\Lambda}$}

For the third model of our study, the system of equations is (\ref{ds1}),
(\ref{ds2}) and%

\begin{equation}
\frac{d%
\Omega
_{\Lambda}}{d\ln a}=-%
\Omega
_{\Lambda}(3%
\Omega
_{\Lambda}-%
\Omega
_{r}-3+n), \label{ds3c}%
\end{equation}

The dynamical system (\ref{ds1}), (\ref{ds2}), (\ref{ds3c}) admits four
critical points, namely $A_{3}=\{1-%
\Omega
_{b},%
\Omega
_{b},0,0\},~B_{3}=\{\frac{n}{3},0,\frac{3-n}{3},0\}$ and $C_{3}%
=\{1,0,0,0\},D_{3}=\{0,0,0,1\}$

Point $A_{3}$ describes a matter (baryons plus dark matter) dominated
universe, hence it that does not accelerate~($w=0$). The eigenvalues of the
linearized system are $\{-1,0,3-n\}.$~For $n<3$ the solution of $A_{3}$ is
always unstable.

Point $B_{3}$\ provides a $\mathbf{\ }\Lambda$CDM scenario where the
components of the fluid are $%
\Omega
_{DM}=\frac{n}{3}~$and $%
\Omega
_{\Lambda}=\frac{3-n}{3}\mathbf{.}$Apparently this family of points exists
only for $0\leq n\leq3$ and it provides cosmic acceleration $(w=\frac{n}%
{3}-1)$ only for $n<2$. The eigenvalues of the critical point are found to be
$\{n-4,n-3,n-3\}$, hence for $n<3$ the point is always unstable. For
$n\rightarrow0$ the solution at the point describes a stable de Sitter
universe $(w=-1)~$where only the cosmological constant term contributes in the
evolution of the universe. Thus this is an interesting point, since
cosmological data point that $n\sim10^{-3}.$
\cite{Gomez-Valent:2014rxa,Sola:2016jky}.

Point $C_{3}$\ describes a dark matter dominated universe that apparently does
not accelerate. The eigenvalues of the linearized system are calculated to be
$\{-1,0,3-n\}$.The point is a source (unstable).

Point $D_{3}$\ describes a radiation dominated universe that does not
accelerate. The eigenvalues of the linearized system are $\{1,1,4-n\}$, from
where we can infer that the solution at point $D_{3}$ is unstable.

The critical point analysis of the above system yields four critical points
that are shown in Table \ref{table3}.~%

\begin{table}[tbp] \centering
\caption{Critical points and physical quantities for Case C}%
\begin{tabular}
[c]{ccccccc}\hline\hline
$\text{Point}$ & $\{%
\Omega
_{DM},%
\Omega
_{b},%
\Omega
_{\Lambda},%
\Omega
_{r}\ \}$ & E$\text{xistence}$ & $w$ & A$\text{cceleration}$ &
E$\text{igenvalues}$ & S$\text{tability}$\\\hline
$A_{3}$ & $\{1-\mathbf{%
\Omega
}_{b},\mathbf{%
\Omega
}_{b},0,0\}$ & Always & $0$ & No & $\{-1,0,3-n\}$ & unstable\\
$B_{3}$ & $\{\frac{n}{3},0,\frac{3-n}{3},0\}$ & $0\leq n\leq3$ & $-1+\frac
{n}{3}$ & Yes for $0\leq n<2$ & $\{n-4,n-3,n-3\}$ & Stable for $n<3$\\
$C_{3}$ & $\{1,0,0,0\}$ & Always & $0$ & No & $\{-1,0,3-n\}$ & unstable\\
$D_{3}$ & $\{0,0,0,1\}$ & Always & $\frac{1}{3}$ & No & $\{1,1,4-n\}$ &
unstable\\\hline\hline
\end{tabular}
\label{table3}%
\end{table}%

\begin{figure}[ptb]
\centering\includegraphics[scale=0.6]{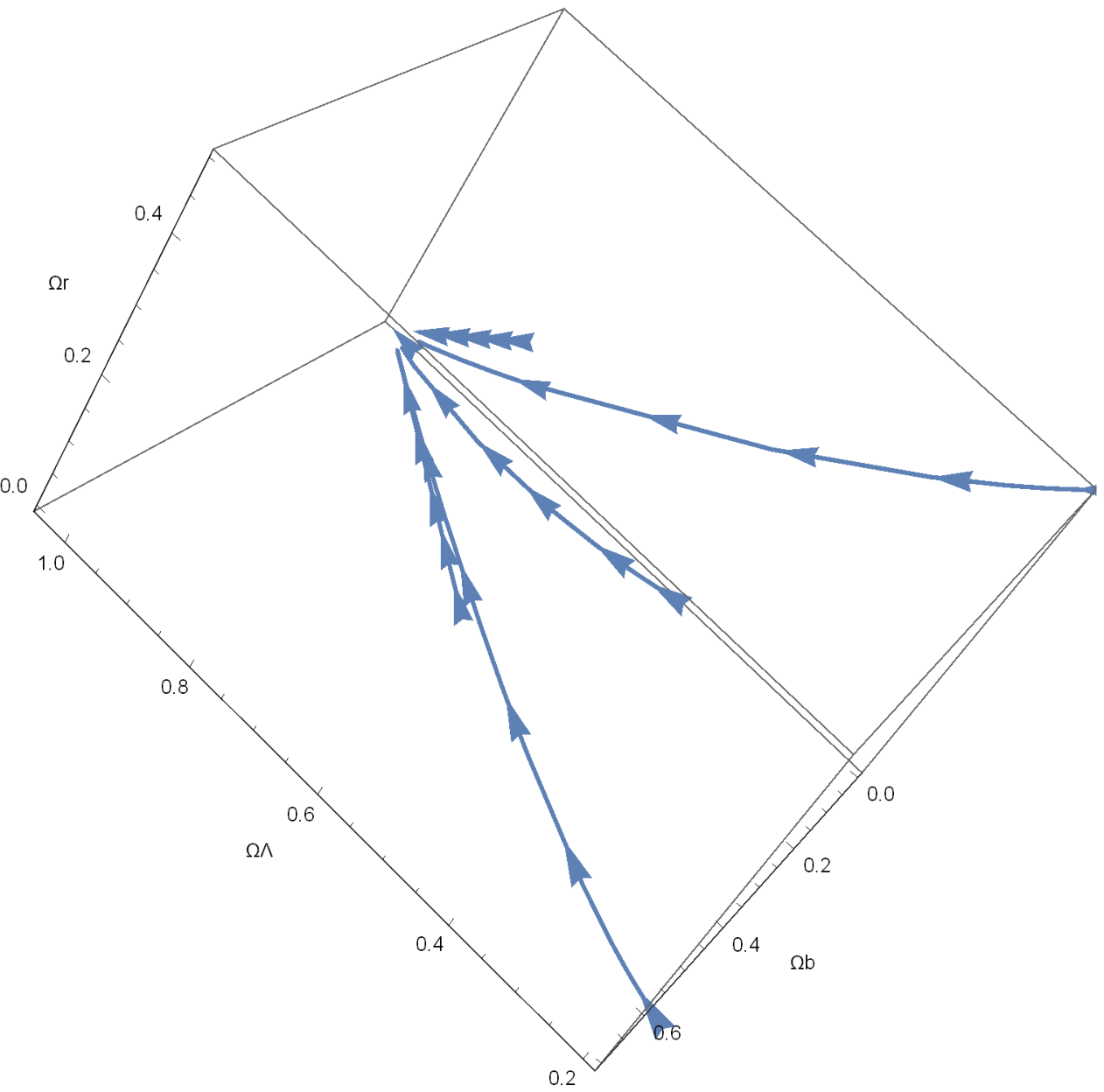}\caption{Phase space diagram
for the dynamical system (\ref{ds1}), (\ref{ds2}), (\ref{ds3c}).We consider
(a) $\Omega_{b}=0.2996,\Omega_{r}=0.0004$,$\Omega_{\Lambda}=0.7$ (b)
$\Omega_{b}=0,\Omega_{r}=0.1$,$\Omega_{\Lambda}=0.9$ (c) $\Omega_{b}%
=0.3$,$\Omega_{r}=0.2$,$\Omega_{\Lambda}=0.5$ (d) $\Omega_{b}=0$,$\Omega
_{r}=0.5$,$\Omega_{\Lambda}=0.2$ (e) $\Omega_{b}=0.7$,$\Omega_{r}=0.1$%
,$\Omega_{\Lambda}=0.2$, for $n<1$. The unique attractor is the point $B_{3}$}%
\label{fig5}%
\end{figure}

\begin{figure}[ptb]
\centering
\includegraphics[scale=0.6]{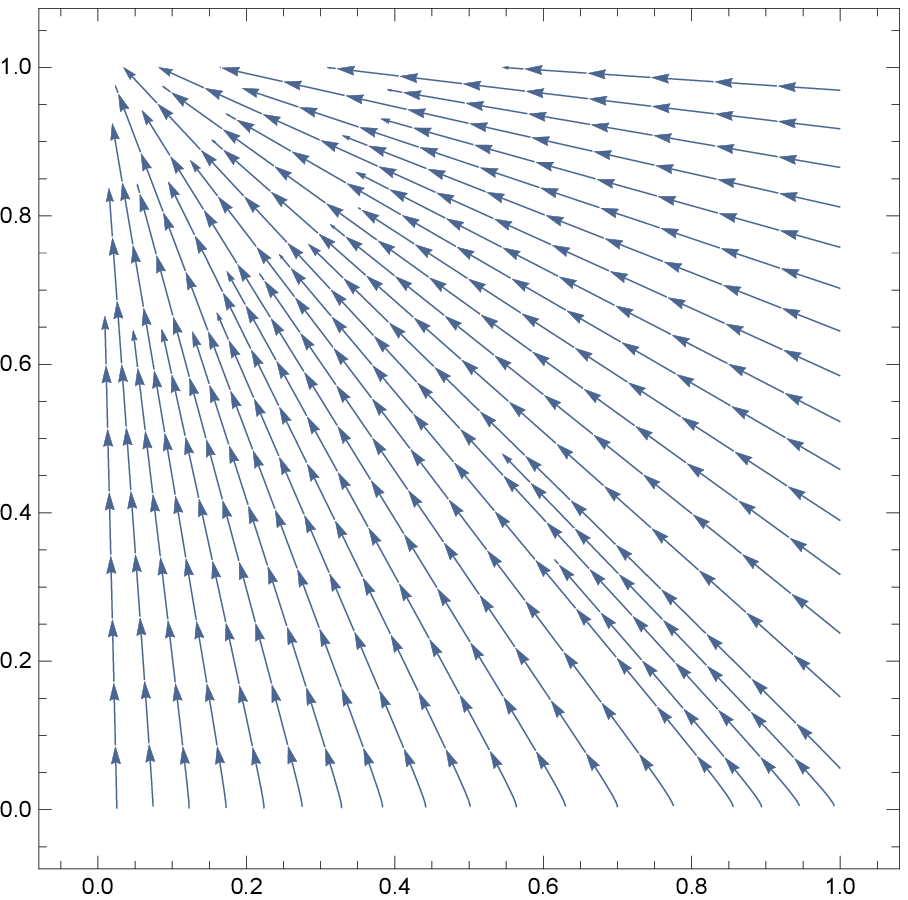}\caption{Phase space diagram for the
dynamical system (\ref{ds1}), (\ref{ds2}), (\ref{ds3c}) in the space of
variables $\Omega_{b},\Omega_{\Lambda}$ for $n<1$ and $\Omega_{r}=10^{-4}$.
The attractor is the point $B_{3}$.}%
\label{fig6}%
\end{figure}

In Figs. \ref{fig5} and \ref{fig6} the phase space diagram of the dynamical
system $Q_{C}$ is presented for $n<1$ \ ($n=-0,1$ ) from where we can see that
the unique attractor is the point $B_{3}$. 

\subsection{Case D - $Q=\frac{3n}{H}\rho_{b}\rho_{DM}$}

For the fourth model of our consideration the dynamical system of our study
consisted by the equations (\ref{ds1}), (\ref{ds2}) and%

\begin{equation}
\frac{d%
\Omega
_{\Lambda}}{d\ln a}=-%
\Omega
_{\Lambda}(3%
\Omega
_{\Lambda}-%
\Omega
_{r}-3)-3n%
\Omega
_{b}(1-%
\Omega
_{b}-%
\Omega
_{\Lambda}-%
\Omega
_{r}),\ \label{ds3d}%
\end{equation}
The dynamical system (\ref{ds1}), (\ref{ds2}), (\ref{ds3d}) admits four
critical points with coordinates $A_{4}=\{0,0,1,0\},~B_{4}=\{1,0,0,0\}$ and
$C_{4}=\{0,1,0,0\},D_{4}=\{0,0,0,1\}.$

Point $A_{4}$ is a viable de Sitter point where only the cosmological constant
term contributes in the evolution of the universe. This point always exists
and it is always stable, since the eigenvalues of the linearized system at
$A_{4}$ are always negative, i.e. $\{-4,-3,-3\}.$

Point $B_{4}$ describes a dark matter dominated universe that does not
accelerate. The eigenvalues are derived $\{3,-1,0\}$ from where we find that
this point is a source.

Point $C_{4}$\ describes a baryon matter only dominated universe that
apparently does not accelerate. The point is a source, because at least one of
the eigenvalues is always positive, the eigenvalues are $\{3,-1,3n\}.$

Point $D_{4}$\ describes a radiation dominated universe that does not
accelerate. The three eigenvaleus are $\{4,1,1\}$, that is, point $D_{4}$ is a
source and the solution described at point $D_{4}$ is unstable.

The critical point analysis of the above system yields four critical points
that are shown in Table \ref{table4}.%

\begin{table}[tbp] \centering
\caption{Critical points and physical quantities for Case D}%
\begin{tabular}
[c]{ccccccc}\hline\hline
$\text{Point}$ & $\{%
\Omega
_{DM},%
\Omega
_{b},%
\Omega
_{\Lambda},%
\Omega
_{r}\ \}$ & E$\text{xistence}$ & $w$ & A$\text{cceleration}$ &
E$\text{igenvalues}$ & S$\text{tability}$\\\hline
$A_{4}$ & $\{0,0,1,0\}$ & Always & $-1$ & Yes & $\{-4,-3,-3\}$ & Stable\\
$B_{4}$ & $\{1,0,0,0\}$ & Always & $0$ & No & $\{3,-1,0\}$ & Unstable\\
$C_{4}$ & $\{0,1,0,0\}$ & Always & $0$ & No & $\{3,-1,3n\}$ & Unstable\\
$D_{4}$ & $\{0,0,0,1\}$ & Always & $\frac{1}{3}$ & No & $\{4,1,1\}$ &
Unstable\\\hline\hline
\end{tabular}
\label{table4}%
\end{table}%

\begin{figure}[ptb]
\centering\includegraphics[scale=0.6]{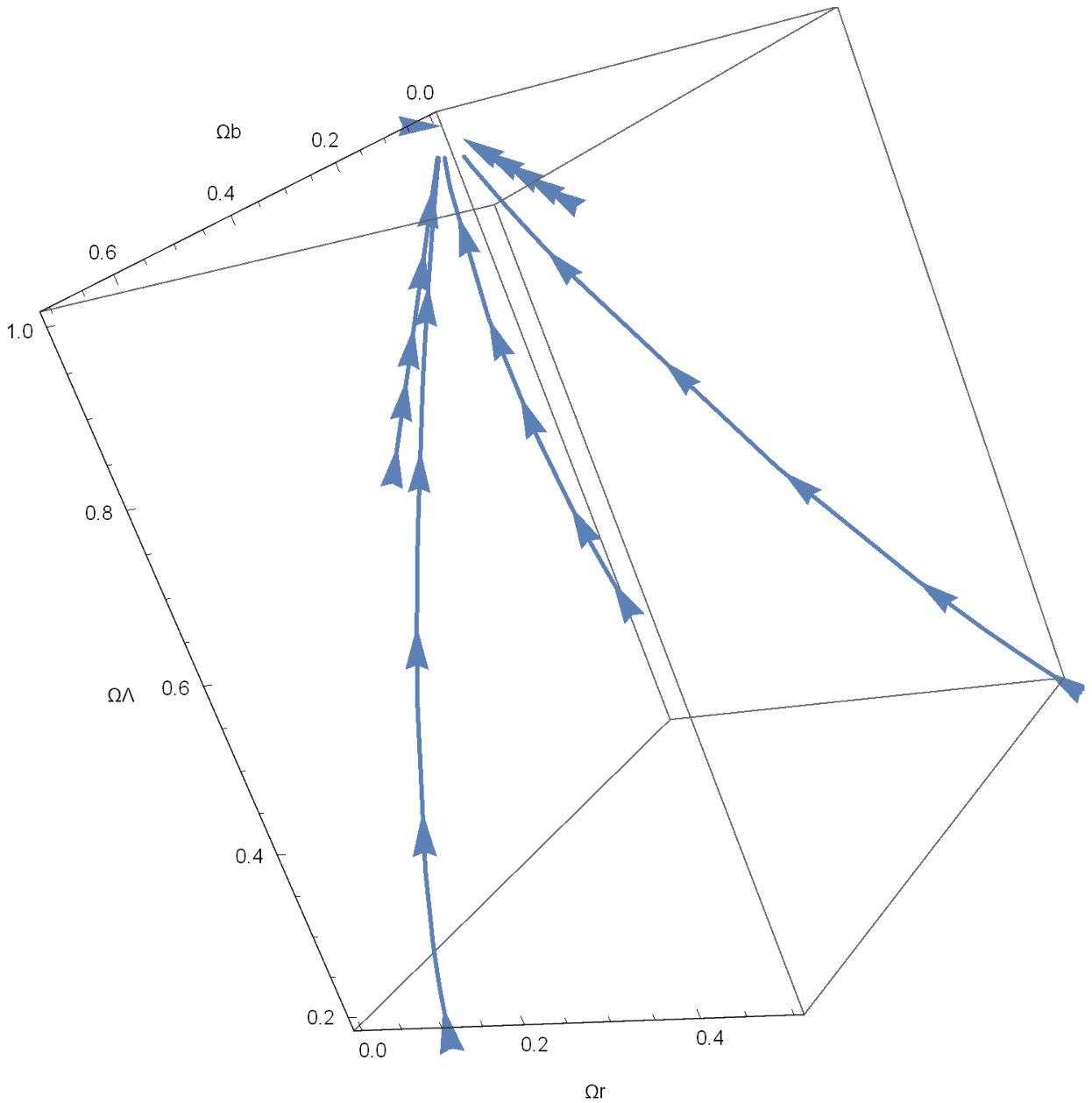}\caption{Phase space diagram
for the dynamical system (\ref{ds1}), (\ref{ds2}), (\ref{ds3d}).We consider
(a) $\Omega_{b}=0.2996,\Omega_{r}=0.0004$,$\Omega_{\Lambda}=0.7$ (b)
$\Omega_{b}=0,\Omega_{r}=0.1$,$\Omega_{\Lambda}=0.9$ (c) $\Omega_{b}%
=0.3$,$\Omega_{r}=0.2$,$\Omega_{\Lambda}=0.5$ (d) $\Omega_{b}=0$,$\Omega
_{r}=0.5$,$\Omega_{\Lambda}=0.2$ (e) $\Omega_{b}=0.7$,$\Omega_{r}=0.1$%
,$\Omega_{\Lambda}=0.2$, for $n<1$. The unique attractor is the de Sitter
point $A_{4}$}%
\label{fig7}%
\end{figure}

\begin{figure}[ptb]
\centering
\includegraphics[scale=0.6]{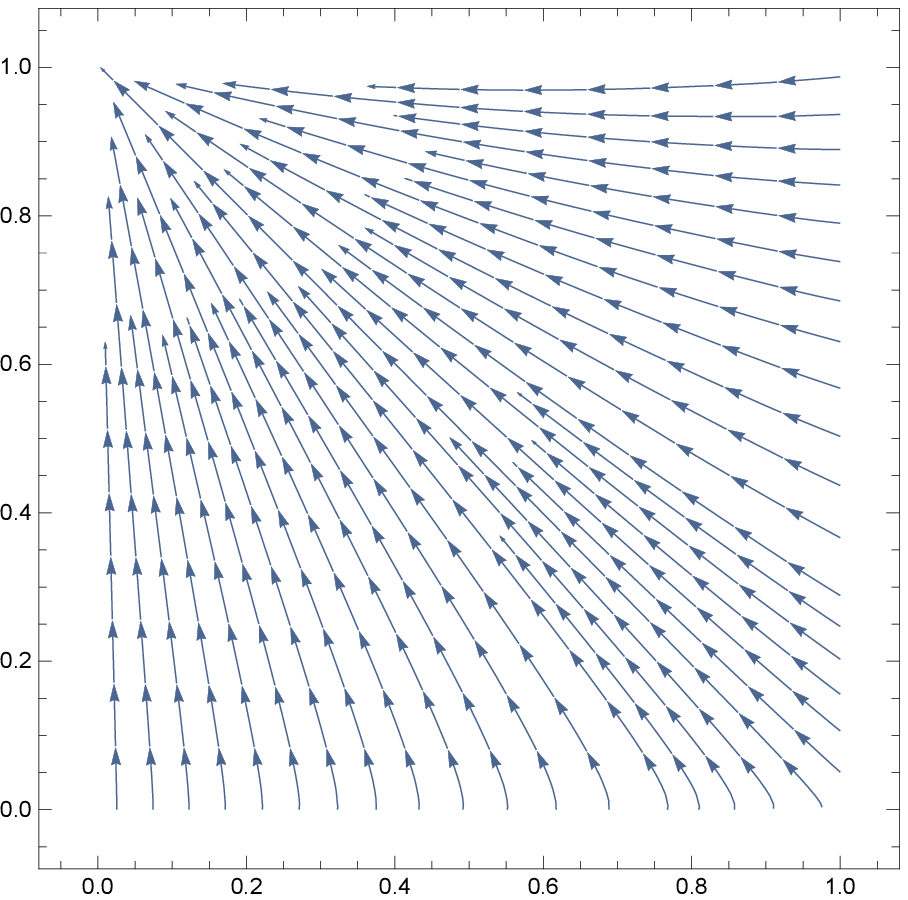}\caption{Phase space diagram for the
dynamical system (\ref{ds1}), (\ref{ds2}), (\ref{ds3d}) in the space of
variables $\Omega_{b},\Omega_{\Lambda}$ for $n<1$ and $\Omega_{r}=10^{-4}$.
The unique attractor is the de Sitter point $A_{4}$.}%
\label{fig8}%
\end{figure}

In Figs. \ref{fig7} and \ref{fig8} the phase space diagram of the dynamical
system $Q_{D}$ is presented for $n<1$ \ ($n=-0,1$ ) from where we can see that
the unique attractor is the de Sitter point $A_{4}$.

\subsection{Case E - $Q=3nH\rho_{tot}$}

For $Q=3nH\rho_{tot}$ the dynamical system of our study consists by the
equations (\ref{ds1}), (\ref{ds2}) and%

\begin{equation}
\frac{d%
\Omega
_{\Lambda}}{d\ln a}=\frac{1}{H}(\frac{\dot{\rho}_{_{\Lambda}}}{3H^{2}}%
-\rho_{\Lambda}\frac{2\dot{H}}{3H^{3}})=-%
\Omega
_{\Lambda}(3%
\Omega
_{\Lambda}-%
\Omega
_{r}-3)-3n, \label{ds3e}%
\end{equation}

The latter dynamical system admits three critical points with coordinates
$A_{5}=\{\frac{1}{2}(1+\sqrt{1-4n},0,\frac{1}{2}(1-\sqrt{1-4n},0\},~B_{5}%
=\{\frac{1}{2}(1-\sqrt{1-4n},0,\frac{1}{2}(1+\sqrt{1-4n},0\}$ and
$C_{5}=\{-3n,0,\frac{3n}{4},\frac{9n+4}{4}\}.$

Points $A_{5}$\ and $B_{5}$ describe both a $\mathbf{\ }\Lambda$-CDM scenario
where the dark matter and the cosmological constant contribute in the
evolution of the universe. Point $A_{5}$ exists for $0\leq n\leq\frac{1}{4}$
and can provide an accelerating universe for $\frac{2}{9}\leq n\leq\frac{1}%
{4}$. Moreover, point $B_{5}$, exists for $0\leq n\leq\frac{1}{4}$ and for the
same range of values can also provide an accelerating universe. As far as the
stability of these two points is concerned, the eigenvalues of the linearized
system at point $A_{5}$ are $\{-\frac{3}{2}(1-\sqrt{1-4n}),-\frac{1}%
{2}(5-3\sqrt{1-4n}),3\sqrt{1-4n}\}_{A_{5}}$, while at point $B_{5}$
are~$\{-\frac{1}{2}(5+3\sqrt{1-4n}),-\frac{3}{2}(1+\sqrt{1-4n}),-3\sqrt
{1-4n}\}$. Therefore, the solution at point $A_{5}$ is always unstable while
point $B_{5}$ is an attractor. \ Furthermore excluding the value $n=\frac
{1}{4}$, in the same range of values it is also a stable point.

Point $C_{5}$ only exists for $n=0$, in which case it describes a radiation
dominated universe $(%
\Omega
_{r}=1)$ that does not accelerate. The eigenvalues are $\{1,\frac{1}%
{2}(5-3\sqrt{1-4n},\frac{1}{2}(5+3\sqrt{1-4n}\}$ which mean that the point is
a source.

The above results are summarized in Tables \ref{table5} and \ref{table6}.%

\begin{table}[tbp] \centering
\caption{Critical points and physical quantities for Case E}%
\begin{tabular}
[c]{|ccccc|}\hline\hline
\multicolumn{1}{||c}{$\text{Point}$} & $\{%
\Omega
_{DM},%
\Omega
_{b},%
\Omega
_{\Lambda},%
\Omega
_{r}\ \}$ & $\text{Existence}$ & $w$ & $\text{Acceleration}$\\\hline\hline
$A_{5}$ & \multicolumn{1}{|c}{$\{\frac{1}{2}(1+\sqrt{1-4n},0,\frac{1}%
{2}(1-\sqrt{1-4n},0\}$} & \multicolumn{1}{|c}{$0\leq n\leq\frac{1}{4}$} &
\multicolumn{1}{|c}{$-\frac{1}{2}(1-\sqrt{1-4n})$} &
\multicolumn{1}{|c|}{$\frac{2}{9}\leq n\leq\frac{1}{4}$}\\\hline
$B_{5}$ & \multicolumn{1}{|c}{$\{\frac{1}{2}(1-\sqrt{1-4n},0,\frac{1}%
{2}(1+\sqrt{1-4n},0\}$} & \multicolumn{1}{|c}{$0\leq n\leq\frac{1}{4}$} &
\multicolumn{1}{|c}{$-\frac{1}{2}(1+\sqrt{1-4n})$} &
\multicolumn{1}{|c|}{$n\leq\frac{1}{4}$}\\\hline
$C_{5}$ & \multicolumn{1}{|c}{$\{-3n,0,\frac{3n}{4},\frac{9n+4}{4}\}$} &
\multicolumn{1}{|c}{$n=0$} & \multicolumn{1}{|c}{$\frac{1}{3}$} &
\multicolumn{1}{|c|}{No}\\\hline
\end{tabular}
\label{table5}%
\end{table}%
%

\begin{table}[tbp] \centering
\caption{Critical points and stability for Case E}%
\begin{tabular}
[c]{|ccc|}\hline\hline
\multicolumn{1}{||c}{$\text{Point}$} & E$\text{igenvalues}$ &
\multicolumn{1}{c||}{S$\text{tability}$}\\\hline\hline
$A_{5}$ & \multicolumn{1}{|c}{$\{-\frac{3}{2}(1-\sqrt{1-4n}),-\frac{1}%
{2}(5-3\sqrt{1-4n}),3\sqrt{1-4n}\}$} & \multicolumn{1}{|c|}{Unstable}\\\hline
$B_{5}$ & \multicolumn{1}{|c}{$\{-\frac{1}{2}(5+3\sqrt{1-4n}),-\frac{3}%
{2}(1+\sqrt{1-4n}),-3\sqrt{1-4n}\}$} & \multicolumn{1}{|c|}{Yes for
$n<\frac{1}{4}$}\\\hline
$C_{5}$ & \multicolumn{1}{|c}{$\{1,\frac{1}{2}(5-3\sqrt{1-4n},\frac{1}%
{2}(5+3\sqrt{1-4n}\}$} & \multicolumn{1}{|c|}{Unstable}\\\hline
\end{tabular}
\label{table6}%
\end{table}%

\begin{figure}[ptb]
\centering\includegraphics[scale=0.6]{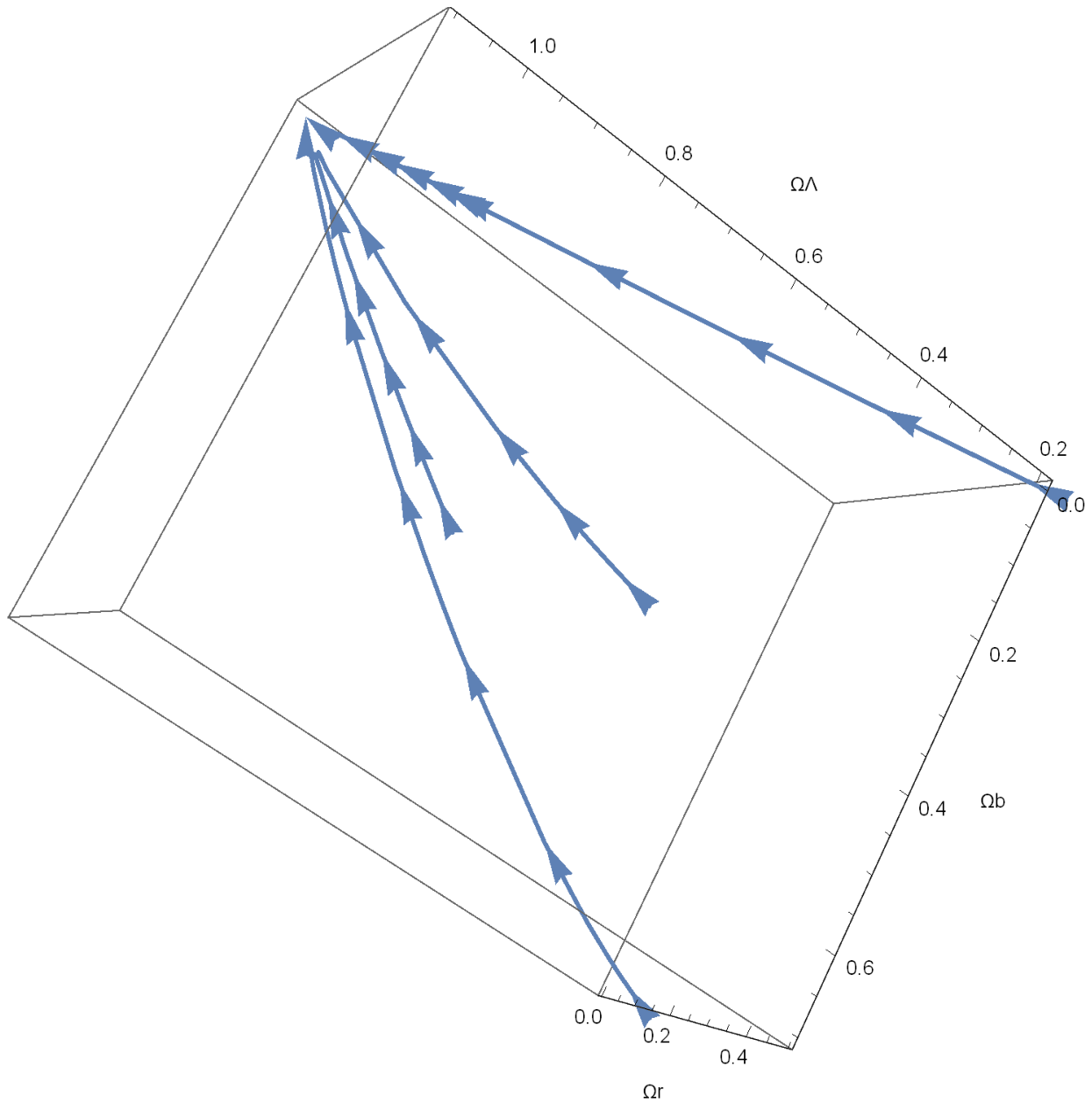}\caption{Phase space diagram
for the dynamical system (\ref{ds1}), (\ref{ds2}), (\ref{ds3e}).We consider
(a) $\Omega_{b}=0.2996,\Omega_{r}=0.0004$,$\Omega_{\Lambda}=0.7$ (b)
$\Omega_{b}=0,\Omega_{r}=0.1$,$\Omega_{\Lambda}=0.9$ (c) $\Omega_{b}%
=0.3$,$\Omega_{r}=0.2$,$\Omega_{\Lambda}=0.5$ (d) $\Omega_{b}=0$,$\Omega
_{r}=0.5$,$\Omega_{\Lambda}=0.2$ (e) $\Omega_{b}=0.7$,$\Omega_{r}=0.1$%
,$\Omega_{\Lambda}=0.2$, for $n<1$. The unique attractor is point $B_{5}$}%
\label{fig9}%
\end{figure}

\begin{figure}[ptb]
\centering
\includegraphics[scale=0.6]{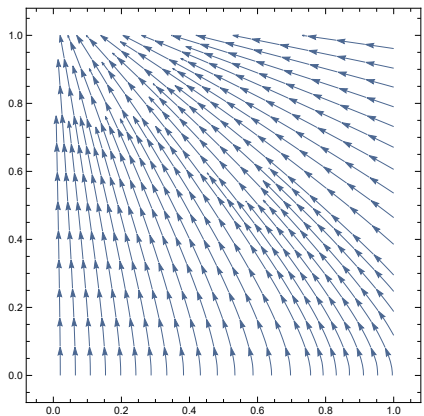}\caption{Phase space diagram for the
dynamical system (\ref{ds1}), (\ref{ds2}), (\ref{ds3e}) in the space of
variables $\Omega_{b},\Omega_{\Lambda}\}$ for $n<1$ and $\Omega_{r}=10^{-4}$.
The unique attractor is the point $B_{5}$.}%
\label{fig10}%
\end{figure}

In Figs. \ref{fig9} and \ref{fig10} the phase space diagram of the dynamical
system $Q_{E}$ is presented for $n<1$ \ ($n=-0,1$ ) from where we can see that
the unique attractor is the point $B_{5}$.

\section{Conclusions}

The current era phenomenology of the $\Lambda-$varying cosmological models has
been discussed by one of the current authors and collaborators, in a number of
very detailed papers. It has been found that the $\Lambda(H)$ models are not
only highly consistent with the plethora of the astrophysical and cosmological
data, but can also help alleviate some of the current-era tensions in data,
including the $\sigma_{8}$ and the current value of the Hubble-parameter
$H_{0}$ tensions\cite{SolaGomezB,GomezSola}. However, a complete dynamical
analysis is missing from the literature. In this article we studied the
dynamical behavior of several varying vacuum models. In particular, we
investigated various models for which baryons and radiation are
self-conserved, while interaction between the dark matter and the varying
vacuum takes different forms. Bellow we summarize the main points of our analysis.

In the first case we assumed the following interaction term $Q_{A}%
=nH(3\rho_{DM}+3\rho_{b}+4\rho_{r}$ from where it follows a viable de Sitter
scenario (point $A_{1}$ as a future attractor for $n<1$). In this scenario $n$
can also have negative values and thus matter is allowed to decay into vacuum.

For our second model, namely $Q_{B}=3nH\rho_{DM}$, we found two possible
interesting scenarios that are described by points $A_{2},B_{2}.$ Point
$A_{2}$ describes again a de Sitter universe that is an attractor for $n<3,$
and point $B_{2}$ describes a $\Lambda$CDM universe that is always unstable
(in the area of its existence $0\leq n\leq3$). This is an interesting result
because this solution recovers $\Lambda$CDM with future attractor an expanding
de Sitter universe.

In the third vacuum model scenario we considered $Q_{C}=3nH\rho_{\Lambda}$,
and found a unique attractor which is described by the critical point $B_{3}$
with $0\leq n<2,$ where the exact solution of this point describes a stable
and accelerating $\Lambda$CDM universe. For the fourth model $Q_{D}=3n\rho
_{b}\rho_{\mathrm{DM}}/H$ a viable de Sitter solution is described by point
$A_{4}$ which is found to be always stable. Finally, for $Q_{E}=3nH\rho
_{\mathrm{tot}}$ we found two points that describe a $\Lambda$CDM universe.
Specifically, point $A_{5}$ with $\frac{2}{9}\leq n\leq\frac{1}{4}$ provides
an unstable $\Lambda$CDM universe, while point $B_{5}$ with $0\leq n\leq
\frac{1}{4}$ provides a stable $\Lambda$CDM model.

It is interesting to mention that in all stable critical points which produce
cosmic acceleration the corresponding parameter $n$ is found to be small,
hence our theoretical results are consistent cosmological observations. Large
values of $n$ lead to a different evolution history for our universe that is
not consistent with the available data. In our analysis, positive values of
$n$ mean that the vacuum decays into dark matter, whereas negative values of
$n$ imply that dark matter decays into vacuum. From our results it is clear
that from the dynamical point of view the interacting varying vacuum scenarios
can largely accommodate models that describe various phases of the observed
behavior of the universe.

\begin{acknowledgments}
GP is supported by the scholarship of the Hellenic Foundation for Research and
Innovation (ELIDEK grant No. 633). SB acknowledges support by the Research
Center for Astronomy of the Academy of Athens in the context of the program
\textquotedblleft Testing general relativity on cosmological
scales\textquotedblright\ (ref. number 200/872). PT acknowledges the support
by the project \textquotedblleft PROTEAS II\textquotedblright\ (MIS 5002515),
which is implemented under the Action \textquotedblleft Reinforcement of the
Research and Innovation Infrastructure,\textquotedblright\ funded by the
Operational Programme \textquotedblleft Competitiveness, Entrepreneurship and
Innovation\textquotedblright\ (NSRF 2014--2020) and co-financed by Greece and
European Union (European Regional Development Fund).
\end{acknowledgments}

\end{document}